\documentclass[preprint,a4paper,onecolumn]{revtex4}

\usepackage[fleqn]{amsmath}
\usepackage{amssymb}
\usepackage[dvips]{graphicx}
\usepackage{color}
\usepackage{tabularx}
\usepackage{algpseudocode}


\begin{document}

\title{Parallel 3-dim fast Fourier transforms with load balancing of the plane waves  }
\author{Xingyu Gao}
\affiliation{Laboratory of Computational Physics, Huayuan Road 6, Beijing 100088, P.R.~China}
\affiliation{Institute of Applied Physics and Computational Mathematics, Fenghao East Road 2, Beijing 100094, P.R.~China}
\affiliation{CAEP Software Center for High Performance Numerical Simulation, Huayuan Road 6, Beijing 100088, P.R.~China}
\author{Zeyao Mo}
\affiliation{Laboratory of Computational Physics, Huayuan Road 6, Beijing 100088, P.R.~China}
\affiliation{Institute of Applied Physics and Computational Mathematics, Fenghao East Road 2, Beijing 100094, P.R.~China}
\affiliation{CAEP Software Center for High Performance Numerical Simulation, Huayuan Road 6, Beijing 100088, P.R.~China}
\author{Jun Fang}
\affiliation{Institute of Applied Physics and Computational Mathematics, Fenghao East Road 2, Beijing 100094, P.R.~China}
\affiliation{CAEP Software Center for High Performance Numerical Simulation, Huayuan Road 6, Beijing 100088, P.R.~China}
\author{Han Wang}
\email{wang_han@iapcm.ac.cn}
\affiliation{Institute of Applied Physics and Computational Mathematics, Fenghao East Road 2, Beijing 100094, P.R.~China}
\affiliation{CAEP Software Center for High Performance Numerical Simulation, Huayuan Road 6, Beijing 100088, P.R.~China}

\begin{abstract}
\noindent
The plane wave method is most widely used for solving the Kohn--Sham equations
in first-principles materials science computations.  
In this procedure, the three-dimensional (3-dim) trial wave functions' fast Fourier transform (FFT) is a regular
operation and one of the most demanding algorithms in terms of 
the scalability on a parallel machine. We propose a new partitioning
algorithm for the 3-dim FFT grid to accomplish the trade-off between the
communication overhead and load balancing of the plane waves. It is shown by
qualitative analysis and numerical results that our approach could scale the plane wave 
first-principles calculations up to more nodes.

\vspace{5pt}
\end{abstract}

\maketitle

\section{Introduction}
\noindent In the context of Density Functional Theory (DFT),
solving the Kohn--Sham equation is the most time-consuming part
of the first-principles materials science computations~\cite{Kohn65, Kresse96CMS, Kresse96PRB}.
The plane wave method, which is a widely used numerical approach~\cite{Payne92},
{could lead} to a large-scale dense algebraic eigenvalue problem. 
This problem is usually solved by the iterative 
diagonalization methods such as Davidson's\cite{Liu1978}, 
RMM-DIIS\cite{Kresse96PRB}, LOBPCG\cite{Knyazev01SIAM}, Chebyshev
polynomial filtering subspace iteration\cite{Zhou06JCP}, etc.
The elementary operation of the iteration methods is the matrix-vector multiplication.
Since the large-scale dense matrix is not suitable for explicit assembly, 
the matrix-vector multiplication is realized
by applying the Hamiltonian operator on trial wave functions. 
{The local term of the effective potential is one part of the Hamiltonian operator.}
In order to compute its action in a lower time complexity, 
we perform 3-dim FFT twice on one trial wave function in each matrix-vector 
multiplication. 

There are three features to make the trial wave function's FFT one of the most
demanding algorithms to scale on a parallel machine. The first is the 
moderate sized FFT grid rather than a large one. The ratio of computation 
to communication of the parallel 3-dim FFT is of order
$\log N$ where $N$, the single dimension of the FFT grid, is usually 
$\mathcal{O}(10^2)$ in most first-principles calculations of bulk materials. 
The second is the accumulated communication overhead led by many execution times corresponding to many wave functions in large-scale problems. 
Thousands of FFTs may run at each step of iterative diagonalization. 
The third is the all-to-all type communication required by the data
transposes. {This can limit the parallel scaling due to the large number of
small messages in the network resulting in competition as well as latency
issues.} 

It has already been recognized that making fewer and larger messages can
speed up parallel trial wave functions' FFTs. 
The hybrid OpenMP/MPI implementation \cite{Goedecker03CPC, Canning2012CSC} can 
lead to fewer and larger messages compared to a pure MPI version. 
And a blocked version \cite{Canning2012CSC} performs a number of trial
wave functions' FFTs at the same time to aggregate the message sizes and reduce
the latency problem.

In first-principles calculations, 
we should consider not only the parallel scaling of wave functions' FFTs, 
but also the load balancing of intensive computations on the plane waves 
that expand the wave functions. The workload of these computations
are usually inhomogeneously distributed on a standard 3-dim FFT grid.
Thus a greedy algorithm is usually used to optimize the load balancing.
However, this algorithm results in global all-to-all communications across all
the processes, thus the latency overhead would grow in proportion to the number 
of processors and might contribute substantially to the total simulation time.
Haynes et.~al.~\cite{Haynes00CPC} present a partitioning approach 
for the 3-dim FFT grid that minimizes the latency cost. Their method depends
critically on the Danielson-Lanczos Lemma \cite{Danielson1942} and requires a particular data
distribution, which limits the possibilities to improve the load balancing of
the plane waves.

\vspace{5pt}
In this paper, we propose a new partitioning method for the 3-dim FFT grid,
  with which we need independent local all-to-all communications for each data transpose rather than
  one global all-to-all communication.
With this communication pattern preserved, we develop the method to improve 
the load balancing by adjusting the data distribution among the working processors.
By numerical examples, we show that
although its load balancing is not as perfect as that of the greedy algorithm,
the new approach can be more favorable for 
parallel scaling by making the fewer and larger messages.
Hence we are allowed to accomplish 
the trade-off between the load balancing of the plane waves and communication 
overhead in the trial wave functions' FFTs. And such a trade-off 
could scale the plane wave first-principles calculations up to more nodes.
With the proposed partitioning method, we design a compact parallel 3-dim FFT to 
reduce the amount of calculations and passing messages without lost of accuracy. 

\vspace{5pt}
The rest of this paper is organized as follows. In Section 2 we introduce
the elemental role of trial wave functions' FFTs in the plane wave method.
In Section 3 we introduce the greedy algorithm for load balancing of the plane waves
and analyze the resulting communication cost. In Section 4 we describe
the new partitioning algorithms and implementations. In Section 5 we show
the numerical results. The last section gives concluding remarks.

\section{Role of trial wave functions' FFT}
In this section, we explain the elemental role of trial wave functions' FFTs
in solving the Kohn--Sham equation using a plane wave basis set.

In the pseudopotential
(norm-conserving \cite{HSC79} or ultrasoft \cite{Vanderbilt90} pseudopotential)
setting or the projector augmented wave (PAW) \cite{Blochl94,Kresse99PRB}
approach, the pseudo wave function $\tilde{\Psi}_i$ satisfies 
the Kohn--Sham equation which looks like:
\begin{equation}
    \left(-\frac{1}{2}\Delta + V_{\mathrm{loc}} + V_{\mathrm{nl}}
    \right)\tilde{\Psi}_i = \epsilon_i S\tilde{\Psi}_i,
    \label{eq:Kohn-Sham}
\end{equation}
where $-\frac{1}{2}\Delta$ is the kinetic energy operator,
$V_{\mathrm{loc}}$ the local potential,
$V_{\mathrm{nl}}$ the nonlocal term,
and $S$ the overlapping operator.
In the case of the norm-conserving pseudopotential,
$S$ could simply be interpreted as the identity operator.
In this paper, we refer to the pseudo wave function simply as the
wave function.

{
We use always the periodic boundary condition and expand the wave functions in
plane waves:
\begin{equation}
    \tilde{\Psi}_{n{\bf k}}({\bf r}) = \sum_{\bf G} \tilde{\Psi}_{n{\bf k}}({\bf G}) \,
e^{-\imath({\bf k}+{\bf G})\cdot{\bf r}},
    \label{eq:pw_expansion}
\end{equation}
where the ${\bf k}$'s are vectors sampling the first Brillouin zone,
$n$ is an index of the energy level with given $k$, 
and ${\bf G}$'s are the reciprocal lattice vectors.} The expansion
\eqref{eq:pw_expansion} only includes the plane waves satisfying
\begin{equation}
    \left|{\bf k}+{\bf G}\right| < \sqrt{2E_{\mathrm{cut}}} \equiv
    G_{\mathrm{cut}}.
    \label{eq:Ecut}
\end{equation}

{
In the plane wave discretization of one large-scale problem,
the Hamiltonian matrix should never be assembled explicitly.
Instead, iterative diagonalization techniques are employed together with the
implicit matrix-vector multiplication that is realized as the action of the
Hamiltonian operator on the trial wave functions.
It is noticed that the local potential is diagonal in the real space.
In order to obtain efficiently the action of the local potential on
the wave function, we should first transform $\tilde{\Psi}_{n{\bf k}}({\bf
G})$ to the real space representation
$\tilde{\Psi}_{n{\bf k}}({\bf r})$ by one FFT,
multiply with the local potential term,
and then transform the product back to the reciprocal space.
Consequently, two 3-dim FFTs are required by each action on a trial wave
function.}

\section{The Load balancing issue and the greedy algorithm}

\subsection{The load balancing issue}\label{sec:layout}

\begin{figure}
    \centering
    \includegraphics[width=0.35\textwidth]{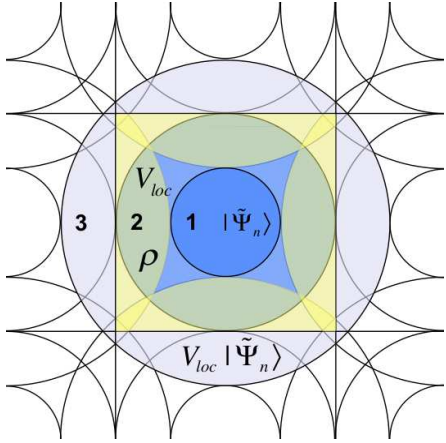}
    \caption{A two dimensional sketch of the wrap-around errors in 
    the periodic reciprocal space. The wave functions $|\Psi\rangle$ 
    is sampled within a sphere with the radius $G_{\mathrm{cut}}$ 
    (the innermost circle 1). The charge density $\rho$
    and the local potential $V_{\mathrm{loc}}$ are
    defined inside a sphere with the radius $2G_{\mathrm{cut}}$ (circle 2).
    We would require a sphere with the radius $3G_{\mathrm{cut}}$ to
    accurately estimate the operation of the local potential on the wave function.
    If we apply a smaller FFT grid with only 
    $2G_{\mathrm{cut}}$, the artificial wrap-around error between 
    $2G_{\mathrm{cut}}$ and $3G_{\mathrm{cut}}$ would occur and be folded back into 
    the 2rd circle due to the periodicity. Hence it is sufficient to
    approximate the wave functions and gradients correctly in circle 1.
    }
    \label{fig:wrap-error}
\end{figure}

As mentioned in the previous section, the plane waves 
are truncated at a certain cut-off radius $G_{\mathrm{cut}}$.
Since the charge density $\rho$ is the sum of squares of the wave functions in the
real space, the corresponding cut-off radius for the charge density is $2G_{\mathrm{cut}}$.
The cut-off radius of the local potential $V_{\mathrm{loc}}$ 
can be regarded the same as that of $\rho$, because $V_{\mathrm{loc}}$ is a
functional of $\rho$. 
Thus, the cut-off radius of $V_{\mathrm{loc}}\tilde{\Psi}_{n{\bf k}}$ is $3G_{\mathrm{cut}}$.
It should be noted that
the Kohn--Sham equation \eqref{eq:Kohn-Sham} is discretized by the
plane waves lying in the sphere of radius
$G_{\mathrm{cut}}$. As illustrated in Fig.~\ref{fig:wrap-error}, it is
sufficient to take the FFT grid with only $2G_{\mathrm{cut}}$ for preventing the wave
functions from the wrap-around error and solving \eqref{eq:Kohn-Sham} correctly.  


\vspace{5pt}
On one hand, we compute the operation of the local potential on the trial wave
functions by the 3-dim FFTs on the standard grid determined by the cut-off
radius $2G_{\mathrm{cut}}$.
{On the other hand, we carry out intensive computations 
time complexities of which are in proportion to the number of the plane waves in a sphere of radius $G_{\mathrm{cut}}$,
including the assembly of the matrix on the subspace, the orthogonalization of wave functions, and
the actions of other parts of the Hamiltonian operator.} Thus the
workload of the intensive calculations is not homogeneously distributed on the grid.
In partitioning the grid, one should consider
not only the parallel scaling of 3-dim FFTs, but also
the load balancing issue of the plane waves calculations.



\subsection{The greedy algorithm}
One 3-dim FFT consists of three successive sets of 1-dim FFTs along the $x$, $y$ and $z$
directions. For each set of 1-dim FFT, the data layout guarantees that each processor holds the complete columns of data along the FFT direction.
Therefore, there are three data layouts of the 1-dim FFTs along the $x$, $y$ and $z$ directions. We call them
the reciprocal space, intermediate and real space layouts, respectively. 

\begin{figure}
    \centering
    \includegraphics[width=0.9\textwidth]{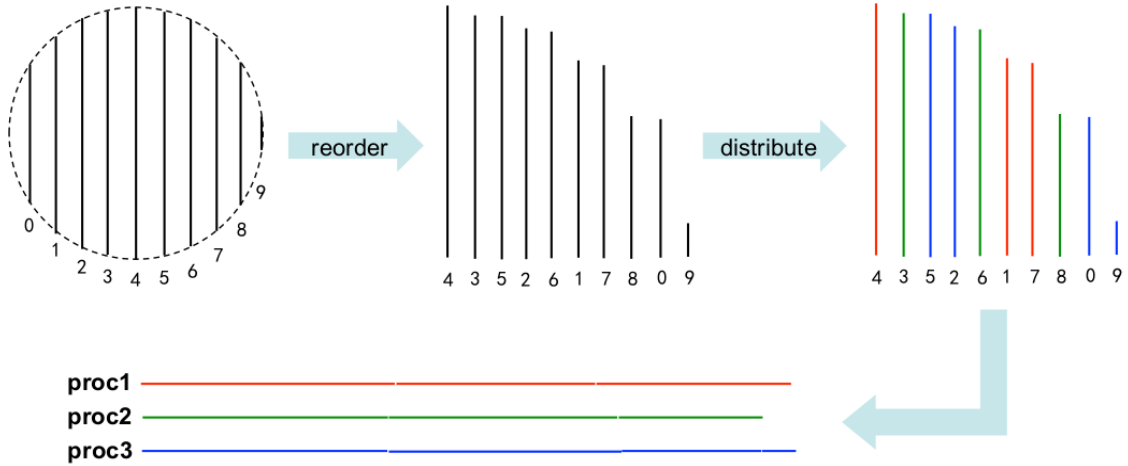}
    \caption{A two dimensional illustration of the greedy algorithm for
    building the reciprocal space layout.}
    \label{fig:greedy-distrib}
\end{figure}

The greedy algorithm is used to build the reciprocal space layout for the sake of load
  balancing. {In the reciprocal space layout, each processor holds the complete
  columns along the $x$ direction.}  
The workload of each complete column is
estimated by the number of plane waves within the
cut-off radius $G_{\mathrm{cut}}$. As illustrated by Fig. \ref{fig:greedy-distrib}, 
we sort these columns in the descending order of workload and
distribute the individual columns among processors in a round robin fashion.
In this way, the reciprocal space layout is established and each processor holds a set 
of complete columns with approximately equal workload.
Due to the uniform distribution of workload, we could directly distribute the
individual columns in a cyclic way to establish the intermediate and real space 
layouts. 


\subsection{The communication pattern and overhead}

When finishing one set of 1-dim FFTs along one direction, 
we transpose the data from the current layout to the next one 
for the successive set of 1-dim FFTs.
The first transpose is between the reciprocal space layout and intermediate
layout, while the second one is between the intermediate layout and real space
layout. With the reciprocal space layout established by the greedy algorithm, 
the first transpose typically requires the all-to-all communication.  
The second transpose may require no communications if each processor 
holds complete planes (perpendicular to the $x$ direction), or limited local communications if each processor has a
section of a plane.

In general, the overhead of the all-to-all data communication mainly consists of two parts: the data transmission and the network latency.
The transmission cost is proportional to the total size of the data packets, and inversely proportional to the internode bandwidth denoted by $\beta$.
The latency cost is proportional to the number of data transmissions initiated.
We denote the latency of one data transmission by $\alpha$.
It worth noting that $\alpha$ and $\beta$ are defined for the situation that a node is sending a data 
packet to another node and simultaneously receiving a packet from another node.

\vspace{5pt}
Without lost of generality, we assume that the all-to-all communications is implemented by the
pairwise data exchanges. Alternative implementations can be found in Ref.~\cite{Rao10}.
Thus the all-to-all communication of $p$ processors can be achieved in 
$p-1$ phases. In each phase, each processor simultaneously sends a data packet
to one processor and receives a packet from another (usually different) processor. 
Though the sizes of data packets are not uniform (in an Alltoallv operation), 
the average size of one packet can be estimated by $\mu N_{\mathrm{FFT}}/p^2$, where 
$\mu$ is the size of a single element (typically 16 bytes for a double precision 
complex data type), and $N_{\mathrm{FFT}}$ is number of the FFT grids.
Hence we estimate the total cost of one all-to-all communication as:
\begin{equation}
    t_1 = (p-1)\left(\alpha+\frac{\mu N_{\mathrm{FFT}}}{\beta p^2}\right).
    \label{eq:comm_t1}
\end{equation}
For a fixed $N_{\mathrm{FFT}}$, the latency overhead grows linearly with respect to the number of
processors, which will probably result in a limited parallel scaling.

\section{The new partitioning algorithm and its implementation}
In this section, we present a new partitioning algorithm of the 3-dim FFT grid
to avoid global all-to-all communications required by the data transposes,
so that the latency cost is alleviated at a cost of small loss of load balancing.



\subsection{The idea of the basic algorithm}\label{sec:basic}

\begin{figure}
    \centering
    \includegraphics[width=0.5\textwidth]{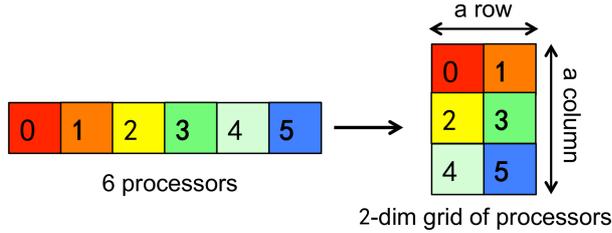}
    \caption{The illustration of a $3\times2$ grid of processors.}
    \label{fig:procs}
\end{figure}

We assume that the number of processors $p$ can be factorized by $m\times n$, where
the difference between $m$ and $n$, i.e.~$|m-n|$, should be as small as possible.
Then the $p$ processors are grouped into $m$ rows by $n$ columns (a $3\times2$ case is illustrated by Fig.~\ref{fig:procs}).
Take the reciprocal space layout for example. We distribute the complete columns
of data along the $x$ direction following two rules: 
1, The data columns with the same $y$-index are distributed within the same column group of processors.
2, The data columns with the same $z$-index are distributed within the same row group of processors.
The intermediate and the real space data layout can be established in a similar way.
The only restrictions are that the intermediate layout 
shares the same data distribution with the reciprocal space layout along the $z$
direction and with the real space layout along the $x$ direction.
In another word, each $xy$ plane in the intermediate layout
  is distributed among the same row of processors as the $xy$ plane in the reciprocal space layout with the same $z$ index, 
  and each $yz$ plane in the intermediate layout
  is distributed among the same column of processors as the $yz$ plane in the real space layout with the same $x$ index.
  
\begin{figure}
    \centering
    \includegraphics[width=0.8\textwidth]{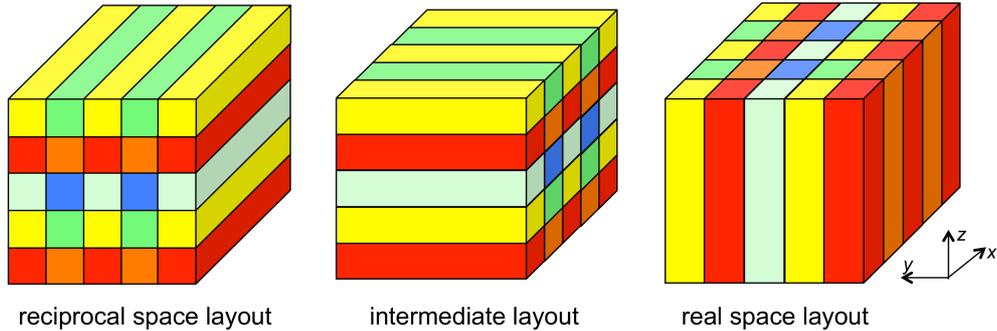}
    \caption{The resulting three layouts by using the basic algorithm
    to partition the 3-dim FFT grid of $5\times5\times5$ among 6 processors.}
    \label{fig:layout_noredis}
\end{figure}

\vspace{5pt}
A direct implementation of the algorithm is to distribute the data columns in
a cyclic fashion. As an illustration, we show how to use it to distribute a 3-dim FFT
grid of size $5\times5\times5$ among 6 processors. The 6 processors 
are grouped into 3 rows by 2 columns, as shown by 
Fig.~\ref{fig:procs}.
The reciprocal space, intermediate and the real space layouts established by our method are shown, from left to right respectively, by
Fig.~\ref{fig:layout_noredis}.
We call this implementation the ``basic algorithm'' in this manuscript.

\vspace{5pt}
In general, the first data transpose (between reciprocal space and intermediate layouts) requires $m$ local all-to-all 
communications which can be carried out \emph{independently} within row groups of $n$ processors.
Similarly, the second transpose (between intermediate and real space layouts) requires $n$ local all-to-all communications 
which can be  carried out \emph{independently} within column groups of $m$ processors.
As shown by Fig.~\ref{fig:layout_noredis}, 
the first data transpose 
requires local all-to-all communications within row groups of two processors,
and the second data transpose requires local all-to-all communications within column groups of three processors.
When $m$ and $n$ are roughly equal to $\sqrt{p}$, the communication overhead can be
estimated as: 
\begin{equation}
    t_2 = (\sqrt{p}-1)\left(\alpha+\frac{\mu N_{\mathrm{FFT}}}{\beta
    p\sqrt{p}}\right).
    \label{eq:comm_t2}
\end{equation}
Compared with the estimated cost \eqref{eq:comm_t1} of the
global all-to-all communication, the growth rate of the latency cost with respect to the number of processors is decreased from $p$ to $\sqrt p$. 

\vspace{5pt}
It should be clarified that both \eqref{eq:comm_t1} and \eqref{eq:comm_t2} are
used to qualitatively illustrate how the communication overhead is decreased 
rather than to give an quantitative interpretation of actual running time. 
Compared with a global all-to-all communication, the new local all-to-all
communications make fewer ($p(\sqrt p - 1)$ v.s.~$p(p-1)$) and
larger ($\mu N_{\mathrm{FFT}}/(p\sqrt p)$ v.s.~$\mu N_{\mathrm{FFT}}/p^2$) messages,
which alleviates the competition as well as  latency issues in the network.
So the proposed partitioning algorithm offers the 
prospect of scaling the plane wave first-principles calculations up to more nodes.

\subsection{The improved algorithm considering the load balancing}\label{sec:adv}
%
The local all-to-all communications can be kept if 
the aforementioned partitioning restrictions are satisfied, i.e. the intermediate layout 
shares the same data distribution with the reciprocal space layout along the $z$
direction, and the same data distribution with the real space layout along the $x$ direction.
So we are allowed to improve the reciprocal space layout considering the load
balancing issue.

{Here we present one possibility:} 
Firstly the workload of each $xz$-plane and $xy$-plane is estimated by the 
number of plane waves in the sphere of radius $G_{\mathrm{cut}}$.
Secondly, the $xz$-planes are sorted in the descending order with respect 
to the workload, and then the reordered $xz$-planes are distributed
to the column groups of processors in a round robin fashion.
Finally, 
the $xy$-planes are sorted in the descending order with respect to the workload,
and then the reordered $xy$-planes are distributed
to the row groups of processors in a round robin fashion.
We will call this implementation the ``improved algorithm'' in this manuscript.


\subsection{The compact 3-dim FFT}

\begin{figure}
    \centering
    \includegraphics[width=0.65\textwidth]{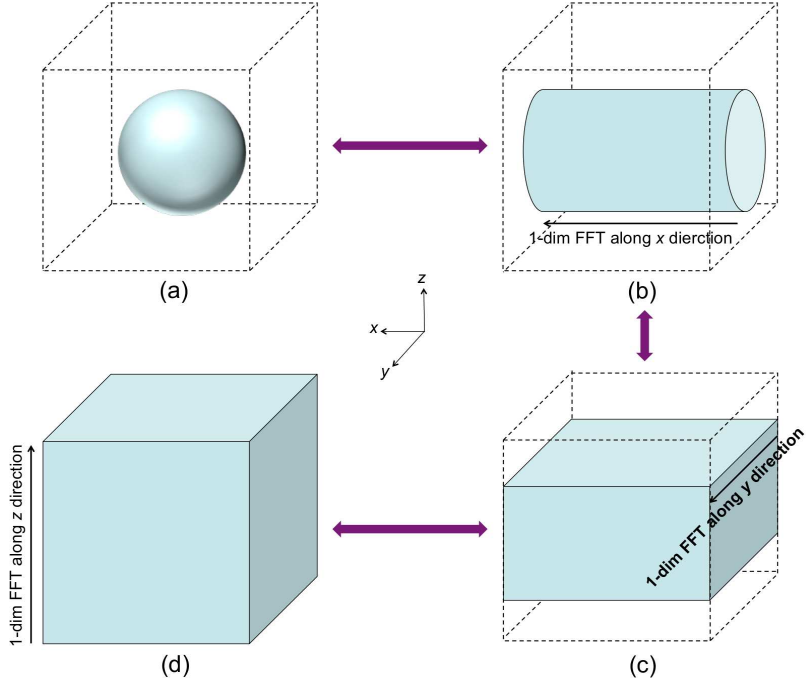}
    \caption{The illustration of a compact 3-dim FFT.
      (a): The aqua region presents the sphere of cut-off radius $G_{\mathrm{cut}}$.
      (b) -- (d):
      The aqua regions present union of all data columns selected to perform the $x$, $y$ and $z$ 1-dim FFTs, respectively.
    }
    \label{fig:compact-FFT}
\end{figure}

As we have discussed in Sec.~\ref{sec:layout},
the standard 3-dim FFT grid is determined by the cut-off radius 
$2G_{\mathrm{cut}}$, while the wave functions are represented by plane waves
within the sphere of radius $G_{\mathrm{cut}}$  (Fig.~\ref{fig:compact-FFT} (a)).
Thus, one can pick up the complete $x$-data columns
that intersect with the sphere to perform 1-dim FFTs along the $x$ direction, 
because all other $x$-data columns contain only vanishing values.
All the selected columns constitute a cylinder, as shown in 
Fig.~\ref{fig:compact-FFT} (b).
After the $x$ direction FFTs, only this cylinder contains non-zero data.
Then one can select the complete $y$-data columns that intersect with
the cylinder to perform the $y$ direction 1-dim FFTs,
and the resulting non-zero-data region is a cuboid, as shown in Fig.~\ref{fig:compact-FFT} (c).
The last set of 1-dim FFTs along the $z$ direction is performed on the whole cube 
shown as in Fig.~\ref{fig:compact-FFT} (d). 
Such a compact 3-dim FFT can also reduce the 
amount of passing messages and calculations compared to the standard 
3-dim FFT implementation that performs 1-dim FFTs on all $x$ and $y$ data
columns in the cube.

If only the $\Gamma$-point is used 
for the ${\bf k}$-point sampling, we can implement a real mode where the
reciprocal space and intermediate layouts can be cut by half since we take
into account that $\tilde{\Psi}_n({\bf G})=\tilde{\Psi}^*_n(-{\bf G})$.


\section{Numerical results}
We implement the parallel compact 3-dim FFT for the trial wave functions 
in the in-house plane wave code package CESSP developed on the infrastructure 
JASMIN \cite{Mo10}. Our implementation is a pure MPI version including 
the greedy algorithm, basic algorithm and improved algorithm.
{With these partitioning algorithms, the parallel scaling of solving the
Kohn--Sham equation \eqref{eq:Kohn-Sham} in PAW approach} are tested on a domestic parallel machine.
Each node of the machine consists of 2
Intel Xeon E5540 CPUs (8 cores) and the nodes are connected by the infiniband
with double data rate (DDR). 

The testing system, {which is sampled by only the $\Gamma$-point, is defined on an 
FCC (face centered cubic) supercell} consisting of 500 Al (aluminum) atoms. 
The self-consistent field iteration runs 7 cycles, and in each cycle the RMM-DIIS algorithm
\cite{Kresse99PRB} is employed to solve the lowest 1001 eigenstates. 
In this process, 38038 FFTs of the trial wave functions are executed one by one. 
The size of the 3-dim FFT grid is $80\times80\times80$,
while the sphere of radius $G_{\mathrm{cut}}$ consists of 35160 plane waves
for the expansion of wave functions.

 
\begin{table}
  \centering
    \caption{Comparison on the parallel scaling of three partitioning algorithms.}
    \label{tab:statlrealF}
    \includegraphics[width=0.95\textwidth]{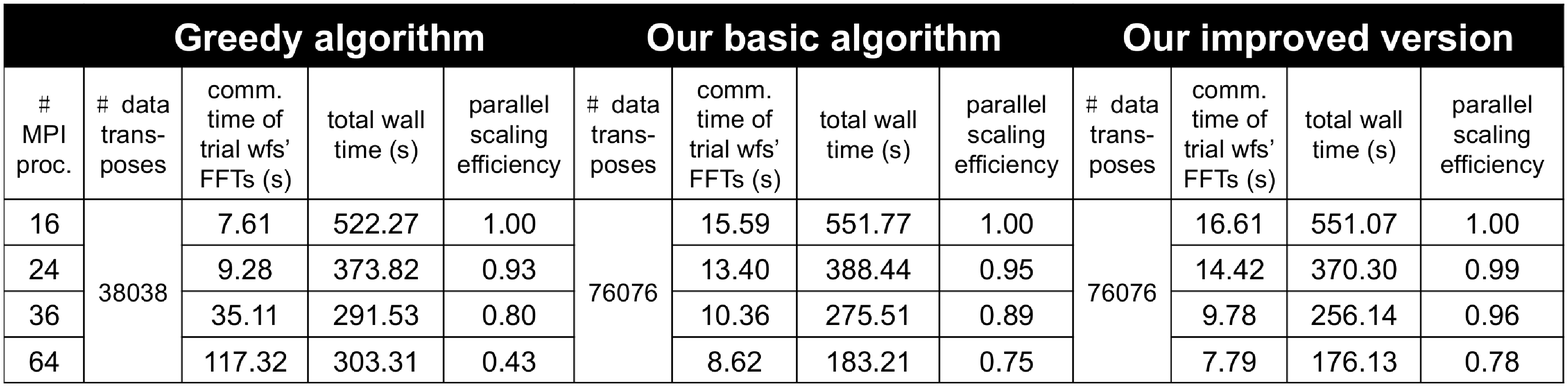}
\end{table}

\vspace{5pt}
In all tests, we launch 8 pure MPI processes per node and count the number 
of data transposes, the total wall time as well as communication time of the 
trial wave functions' FFTs. The results of the tests are summarized in Tab.~\ref{tab:statlrealF}.
In the greedy algorithm, no data transposes are required between the
intermediate and real space layouts since the intermediate layout holds the
complete $yz$-planes on each processor. So the greedy algorithm needs half
number of data transposes as our new algorithms.  
Nevertheless, as shown by Tab.~\ref{tab:statlrealF}, with increasing number of 
processors, the greedy algorithm leads to a rapid growth in the communication cost,
which finally takes more than one third of the total computational cost, while 
the new algorithms (both the basic and its improved versions) can 
effectively suppress the growth in the communication cost.
When the number of processors is less than 24, the greedy algorithm is preferable, while
when the number of processors is more than 24, our algorithms could provide better overall performance.

\begin{table}
    \centering
    \caption{Comparison on the load balancing of three partitioning
    algorithms}
    \label{fig:loadbalancing}
    \includegraphics[width=0.95\textwidth]{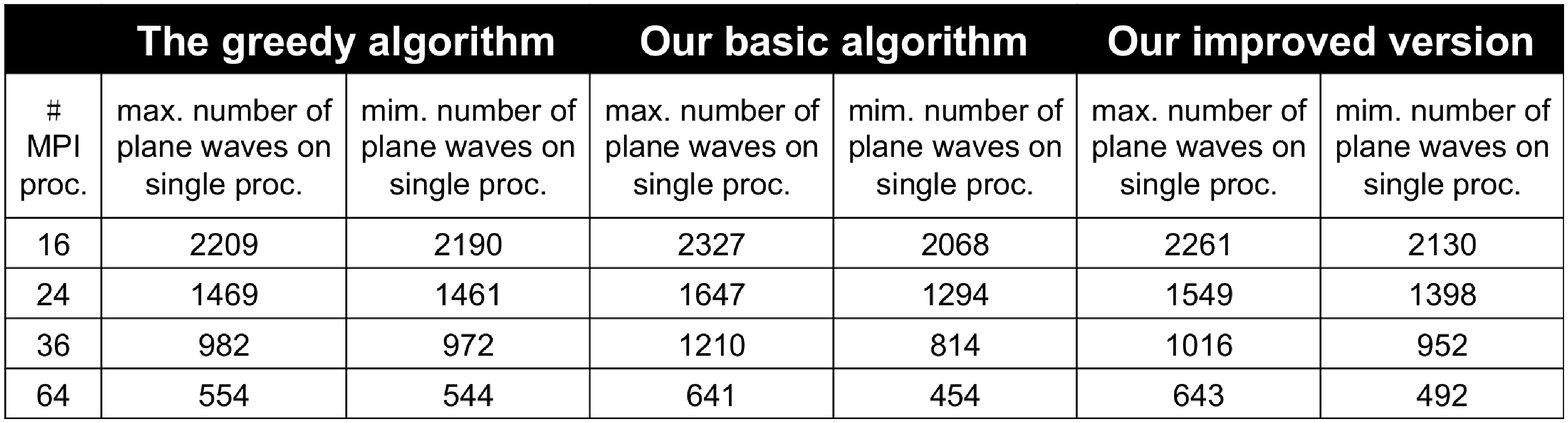}
\end{table}

In Tab.~\ref{fig:loadbalancing} we represent the load balancing in the simulations by comparing the maximum
and minimum numbers of plane waves distributed on a single processor.
It is obvious that the load balancing of the greedy algorithm is almost perfect.
{Although not as perfect as the greedy algorithm, the load balancing
of our improved algorithm is still acceptable.}
Comparing with the basic algorithm,
the improved version can effectively reduce the gap between the maximum and
minimum number of plane waves on a processor.
Combining Tab.~\ref{tab:statlrealF} and \ref{fig:loadbalancing} together, we
present an example of { 
the trade-off between the load balancing and communication cost:
the greedy algorithm has best load balancing but could lead to very limited parallel
scaling; our new algorithms would achieve much better scaling at a moderate loss of load balancing.}



\section{Conclusion}
We present a new partitioning algorithm for the 3-dim FFT grid used in 
the plane wave first-principles calculations. Compared with the greedy algorithm 
{biased toward} load balancing of the plane wave computations,
our approach primarily {suppresses the growth in communication overhead {with respect to increasing number of processors} by
performing local all-to-all communications during data transposes. 
{Then we adjust the} data distribution to improve the load
balancing with the communication pattern preserved.
In the numerical examples, we present a trade-off:
a much lower communication overhead on a relatively large number of
processors is achieved at a moderate loss of load balancing.
By using the new algorithm, we could scale the whole plane wave codes up to more processors than the greedy algorithm.
For a better performance, our approach can be seamlessly combined with 
other techniques such as the hybrid OpenMP/MPI implementation or simultaneously
performing a large number of FFTs.


\section*{Acknowledgment}
The authors would like to thank Hong Guo and Zhang Yang for useful
discussions. And the first author is especially thankful to Xiaowen Xu for his
encouragement.
This work was partially supported by the National Science Foundation of China under
Grants 91430218 and 61300012, and the National High Technology Research and Development 
Program of China under Grant 2015AA01A304.
H.W. is supported by the National Science Foundation of China under
Grants 11501039 and 91530322.


\end{document}